# Reply to Comment on: "Spontaneous breaking of time-reversal symmetry in the pseudogap state of a high-T*c* superconductor"

A. Kaminski *†, S. Rosenkranz *†, H. M. Fretwell ‡, J. C. Campuzano *†, Z.Z. Li §, H. Raffy §, W. G. Cullen †, H. You †, C. G. Olson ||, C. M. Varma ¶, and H. Höchst #

*Department of Physics, University of Illinois at Chicago, Chicago, Illinois 60607, USA
†Materials Science Division, Argonne National Laboratory, Argonne, Illinois 60439,USA
‡Department of Physics, University of Wales Swansea, Swansea, UK
§Laboratoire de Physique des Solides, Université Paris-Sud,91405 Orsay, France
||Ames Laboratory, Iowa State University, Ames, Iowa 50011, USA
¶Bell Laboratories, Lucent Technologies, Murray Hill, New Jersey, 07974, USA
#Synchrotron Radiation Center, Stoughton, Wisconsin 53589, USA

In a recent comment [1], Armitage and Hu have suggested that our experiment observing dichroism in angle resolved photoemission (ARPES) [2] could not be conclusively interpreted as arising from time reversal symmetry breaking, arguing that our observations are likely due to structural effects. The concerns expressed by Armitage and Hu that our results could be due to a change in the mirror plane are as important as they are obvious. In fact the first part of their comment merely restates the results of Simon and Varma [3] about the relationship and contrast of effects due to time reversal symmetry breaking and those caused by crystallographic changes. In any test of time reversal symmetry one must ensure that parity alone is not inducing the observed changes. We have indeed considered this issue very carefully in the course of our study [2] and it is precisely the lack of temperature dependent structural changes significant enough to explain the magnitude of the observed dichroism that forced us to conclude that time reversal symmetry breaking is the only plausible explanation.

The effect of the superstructure modulation (SSM) on the ARPES signal was studied in detail and is well understood [4]. The SSM -- a buckling of the BiO layer -- acts as a diffraction grating on the photoelectrons, producing "ghost" images of the Fermi surface.
There are two issues to consider here: (a) magnitude of the dichroism signal as compared to magnitude of "contamination" signal arising from "ghost" images due to SSM, (b) the temperature dependence of both effects. (a) Along the $(\pi,-\pi)=>(\pi,\pi)$ direction, the first "ghost" image of the Fermi surface extends from $(\pi, -0.39)$ to $(\pi, -0.03)$ and the second from $(\pi,0.03)$ to $(\pi, 0.39)$, as shown in Fig. 1. Therefore, $(\pi,0)$, which was the main focus of our investigation, being outside of that range, is not significantly contaminated by the signal from the superstructure modulation. Furthermore the sign of the dichroism in the immediate surroundings of $(\pi,0)$ is opposite for the two ghost images due to the anti-symmetry of the geometric effect about the $(0,0)=>(\pi,\pi)$ axis. Any changes in the dichroism due to the SSM are therefore very small, as the contributions from the two SSM images cancel each other. It is then easy to understand that a very large change of several times the SSM intensity would be required to account for the observed dichroism of 3%. In addition, the plots of the dichroism signal D in Ref. [2] are straight lines. It can be seen from Fig. 1, that if the SSM contribution were to be large, D would significantly deviate from linearity near k=+-0.05.
(b) There is no temperature dependence to the SSM contribution, as we show in Fig. 2, where we plot n(k) over an extended cut along the diagonal direction of the Brillouin zone. The only observed change in n(k) with temperature is the expected broadening.

Another possible structural effect suggested by Armitage et al [1], and discussed in Ref. [2], could be a rotation of the mirror plane due to a change in the orthorhombicity of the lattice. A rotation necessary to account for the magnitude of the dichroism seen in our experiments would be close to 2°. This of course would be easily observable even in a rudimentary Laue diffraction pattern. However, the work of Miles et al. [5] cited by Armitage and Hu measures the total change in the b-axis lattice parameter between 50K and 300K to be 0.011 Å. Even assuming a non-expanding a-axis, this would correspond to a rotation of the crystal symmetry axis of 0.05°, or two orders of magnitude smaller than what is neeccesary to account for the dichroism effect observed in our experiment. Our own x-ray diffraction measurements (shown in Fig. 3) performed on the same samples as used in the dichroism experiments are in good agreement with the data by Miles et al. [5] Furthermore, we examined eight diffraction peaks, all equivalent (1,0,5) peaks in the pseudotetragonal notation,i.e. (1,0,5), (-1,0,5), (0,1,5), (0,-1,5) and all four (1,1,10)'s, below and above T*, yielding consistent results.

Any structural effects can be detected through the change in symmetry in ordinary, linearly polarized ARPES. This has the additional advantage that the "dichroic" and the "ordinary" ARPES are both surface sensitive experiments. The fact that the mirror plane does not rotate to any significant degree is indeed seen in the ARPES measurement of the dispersion (using linearly polarized photons), which retains its even symmetry to a high degree of precision, as shown in Fig. 4.

Figure captions.

Fig. 1 Schematic band structure and SSM diffraction images along the $(\pi,-\pi)-(\pi,\pi)$ direction (marked as thick black line on the inset) adopted from Ref. [5].

Fig. 2 ARPES data along the diagonal direction for T=50K and T=200K for an underdoped sample and integrated intensity (-600 meV, +100 meV) showing no significant changes of the SSM signal as function of temperature

Fig. 3 Upper limit on the rotation of the mirror plane derived from x-ray scattering on the same underdoped sample as used in Ref. [2].

Fig. 4 Band dispersion along the line marked in the inset obtained from MDC fits to ARPES data for underdoped sample in the pseudogap state showing high degree of symmetry.

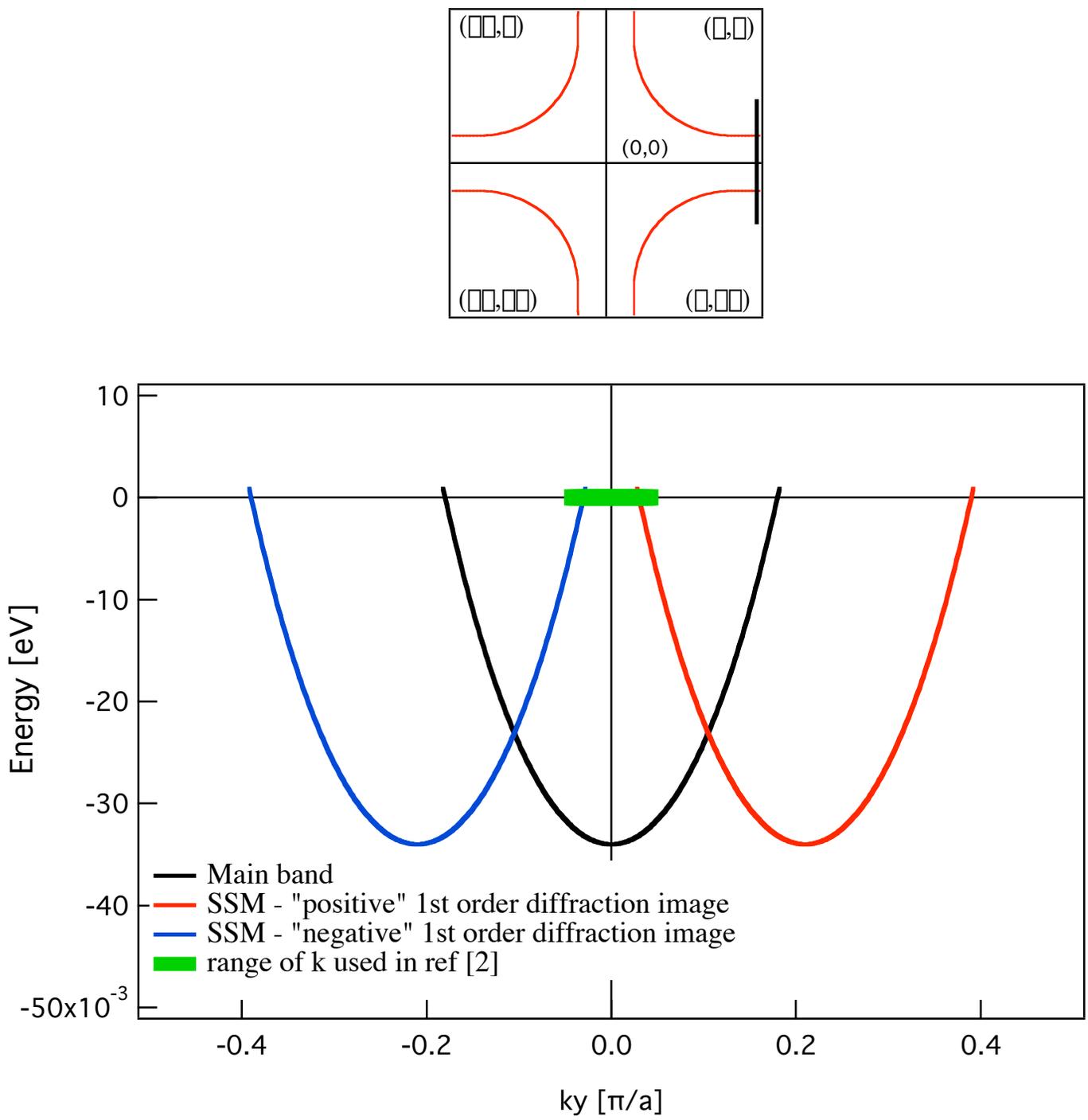

Fig. 1

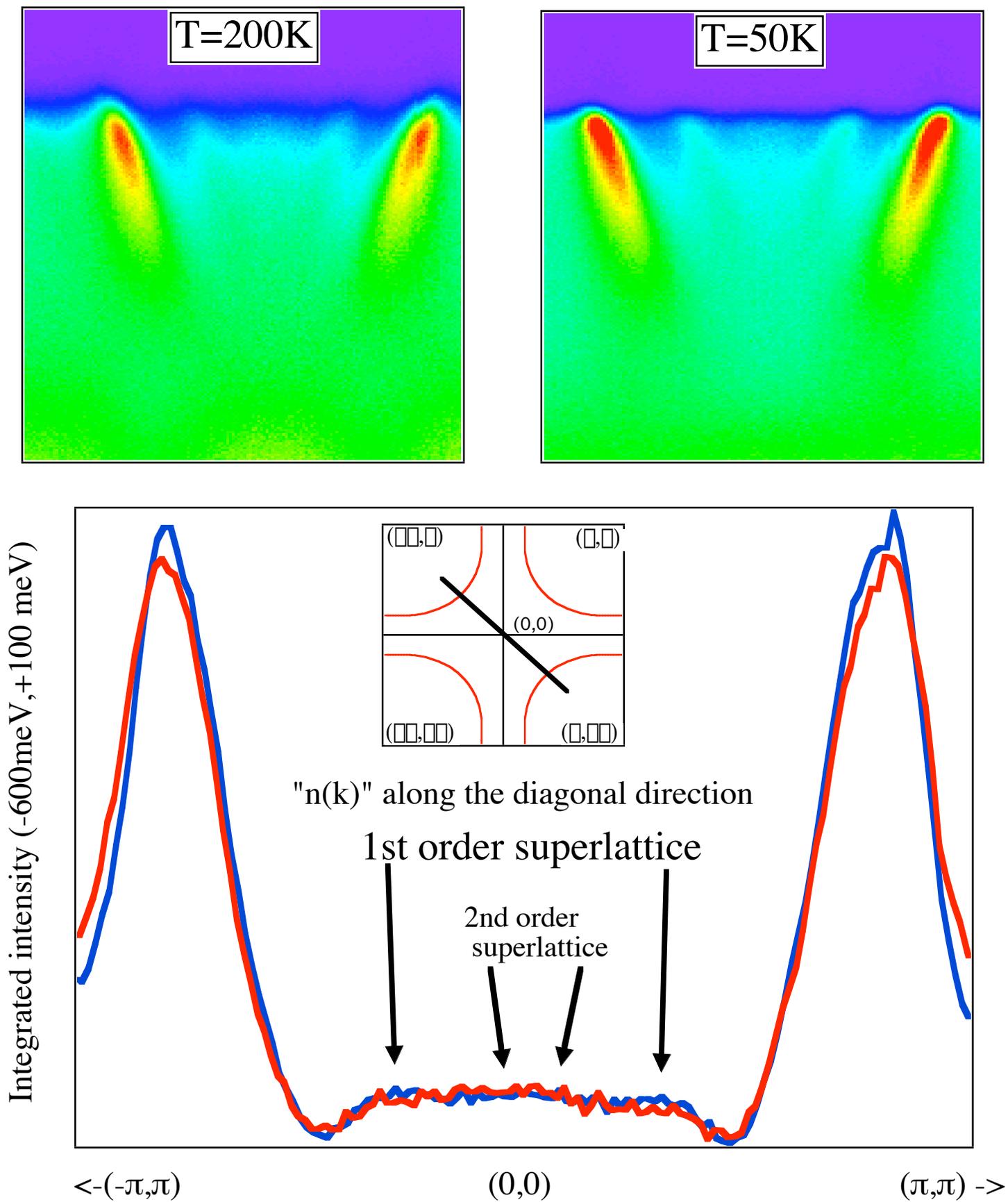

Fig. 2

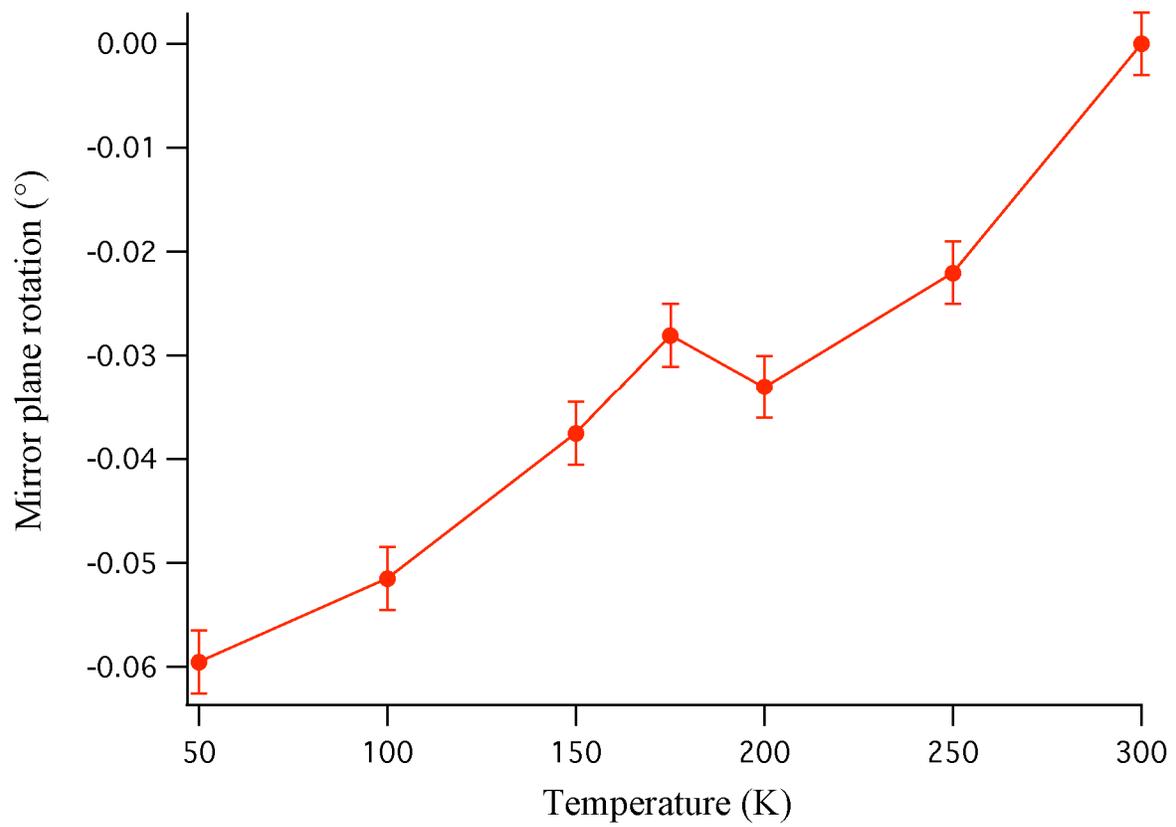

Fig. 3

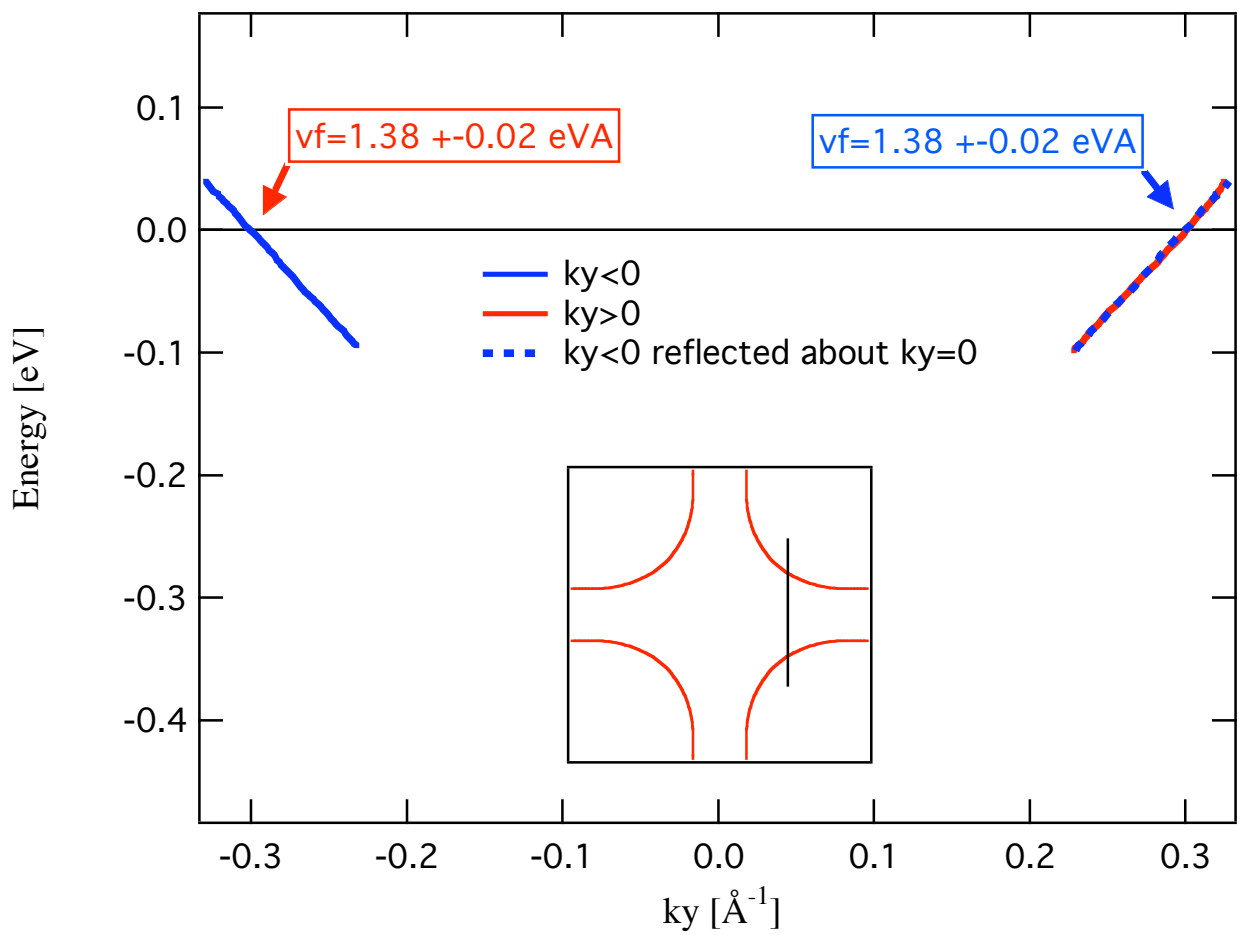

Fig. 4